\documentstyle[a4,psfig]{article}
\begin{document}

\title{Probability distributions of the work\\ in the 2D-Ising model}

\author{C. Chatelain and D. Karevski\\
Groupe M,\\
Laboratoire de Physique des Mat\'eriaux,\\
Universit\'e Henri Poincar\'e Nancy I,\\
BP~239, Boulevard des aiguillettes,\\
F-54506 Vand{\oe}uvre l\`es Nancy Cedex, France\\
{\small\tt chatelai@lpm.u-nancy.fr, karevski@lpm.u-nancy.fr}}

\maketitle

\begin{abstract}
Probability distributions of the magnetic work are computed for the 2D Ising model
by means of Monte Carlo simulations. The system is first prepared at equilibrium for
three temperatures below, at and above the critical point. A magnetic field is then
applied and grown linearly at different rates. Probability distributions of the work
are stored and free energy differences computed using the Jarzynski equality.
Consistency is checked and the dynamics of the system is analyzed. Free energies
and dissipated works are reproduced with simple models. The critical
exponent $\delta$ is estimated in an usual manner.
\end{abstract}


\def\build#1_#2^#3{\mathrel{
\mathop{\kern 0pt#1}\limits_{#2}^{#3}}}
\def\ket#1{\left| #1 \right\rangle}
\def\bra#1{\left\langle #1\right|}
\def\braket#1#2{\left\langle\vphantom{#1#2} #1\right.
\left| \vphantom{#1#2}#2\right\rangle}
\def\spring{\hskip 0pt minus 1fil}
\def\identite{{\rm 1}\hbox to 1 pt{\spring\rm l}}
\def\trace{\mathop{\rm Tr}\nolimits}

\section{Introduction}
The Jarzynski equality~\cite{Jarzynski97a,Jarzynski97b} is one of the few exact
results in the context of out-of-equilibrium statistical physics. This simple and
elegant relation applies to systems initially prepared at thermal equilibrium and
then driven out-of-equilibrium by varying a control parameter $h$ from say $h_1$
to $h_2$. The probability distribution of the work $W$ extracted during the
experiment is related to the free-energy difference $\Delta F=F(h_2)-F(h_1)$
between the two equilibrium states at values $h_1$ and $h_2$ of the control
parameter by:
	\begin{equation}
	e^{-\beta\Delta F}=\langle e^{-\beta W}\rangle.
	\label{eq1}
	\end{equation}
Remarkably, this relation gives some information about the
equilibrium state at the value $h_2$ of the control parameter even though this
state has never been reached by the system. For a cyclic transformation, one recovers
the Bochkov-Kuzovlev relation $\langle e^{-\beta W}\rangle_{\rm Cycl.}=1$~\cite{Bochkov81}.
As recognized by Crooks, the Jarzynski equality can be derived from a more general
fluctuation theorem~\cite{Crooks99}. It  has passed experimental, see for
instance~\cite{Liphardt02,Ritort03,Douarche05} as well as numerical~\cite{Lua05} tests.
In both cases, the importance of a sufficiently accurate sampling of the tail
of the probability distribution $\wp(W)$ has been emphasized. Since the pioneering
experiment on DNA by Liphardt {\sl et al.}~\cite{Liphardt02}, the Jarzynski
relation is now widely used in chemistry and biophysics to estimate equilibrium
free energy differences based on experiments or short-time out-of-equilibrium
numerical calculations. However, the way the interaction with the thermal
bath is taken into account in the original Jarzynski's demonstration has been
criticized~\cite{Cohen04}: the interaction is assumed to be weak enough
to be neglected. It means that the system cannot exchange heat with the
bath. Equation (\ref{eq1}) was thus claimed to fail since one expects that the
system tends to relax to an equilibrium state by exchanging heat with the bath.
For Jarzynski's response, see~\cite{Jarzynski04}. As far as
we know, no experiment has provided any evidence of this failure yet. In this paper,
we present a Monte Carlo study of the 2D Ising model with a Glauber dynamics.
For such Markovian dynamics, the interaction with the bath is properly taken
into account by the transition rates. The criticisms do not apply to this case
and the demonstration given by Jarzynski in appendix of
reference~\cite{Jarzynski97b} is exact.
\\

In the context of spin models, up to now few studies have taken advantage of
the Jarzynski equality to calculate the free energy. Results have recently been
obtained for a single Ising spin~\cite{Marathe05} and in the mean-field
approximation~\cite{Imparato05}. We present a Monte Carlo investigation of the
two-dimensional Ising model. Numerical details of the calculation are
presented in the first section. The three next sections correspond to calculations
with an initial equilibrium state at different temperatures:
in the paramagnetic phase ($\beta=0.2$), the ferromagnetic phase ($\beta=0.7$)
and at the critical point ($\beta\simeq 0.4407$).

\section{Numerical procedure}
We study the two-dimensional Ising model defined by the usual Hamiltonian
	\begin{equation}
	{\cal H}(\{\sigma\})=-J\sum_{(i,j)} \sigma_i\sigma_j-h\sum_i \sigma_i,
	\quad\sigma_i=\pm 1
	\end{equation}
where the sum extends over nearest-neighbors on a square lattice. In the following,
the exchange coupling is fixed to $J=1$. Lattices with sizes from $32\times 32$ to
$128\times 128$ and periodic boundary conditions have been considered. The
system is first prepared in an equilibrium state without magnetic field using
the Swendsen-Wang cluster-algorithm~\cite{SwendsenWang87}. In the paramagnetic
and ferromagnetic phases, the system was first equilibrated using 200 Monte
Carlo Steps (MCS). At the critical point, we used 1000 MCS to circumvent the
critical slowing-down. This last value is a safe bet since it is more than two
orders of magnitude larger than the autocorrelation time $\tau_E\simeq 5.87(1)$
at $T_c$ and for $L=128$~\cite{Salas96}. We then let the system evolve
according to a local dynamics, i.e. using the Metropolis
algorithm~\cite{Metropolis53}, during $n_{\rm iter.}$ Monte Carlo iterations
while the magnetic field $h$ is linearly increased. In the following, $h$
denotes the final magnetic field reached after $n_{\rm iter.}$ iterations.
The work is calculated as 
	\begin{equation}
	W=-\int_0^h Mdh=-\int_0^{t_f} M(t)\dot h dt
	\simeq -{h\over n_{\rm iter.}}\sum_{i=0}^{n_{\rm iter.}-1} M_i
	\end{equation}
i.e. as the average magnetization during these $n_{\rm iter.}$ iterations times
the magnetic field $h$. Note that $W$ is the mechanical work and not the work as
usually defined in thermodynamics~\cite{Narayan03}. Let us emphasize that the
demonstration of (\ref{eq1}) given by Jarzynski at the end of his second paper on
the subject~\cite{Jarzynski97b} requires a well-defined protocol: first a sudden
change $\Delta h$ of the magnetic field is performed leading to a work $-\Delta
hM_i$ where $M_i$ is the magnetization and second a Monte Carlo iteration is
made which allows the system to relax and exchange heat $Q=\Delta E=E_{i+1}-E_i$
with the bath. Note that when $n_{\rm iter.}=1$, the Jarzynski relation is
equivalent to the thermodynamic perturbation. The experiment is repeated $n_{\rm
exp.}=100,000$ times. Instead of restarting the whole simulation, we used the
last spin configuration obtained without magnetic field and did 10 further
Swendsen-Wang MCS at $T\ne T_c$ and 50 at $T_c$. This last value is almost ten
times larger than the autocorrelation time. Autocorrelations between two
successive experiments are thus smaller than ${\rm exp}(-50/5.9)\simeq
2.10^{-4}$ and will be neglected in the following. Averages are computed as
	\begin{equation}
	\langle f(W)\rangle=\int f(W)\wp(W)dW
	={1\over n_{\rm exp.}}\sum_{\alpha=1}^{n_{\rm exp.}} f(W_\alpha)
	\end{equation}
where $\{W_\alpha\}_\alpha$ is the set of values obtained when repeating $n_{\rm
exp.}$ times the numerical experiment. Errors are estimated as
$\sqrt{[\langle f^2\rangle-\langle f\rangle^2]/n_{\rm exp.}}$ as expected from
the central limit theorem for uncorrelated random variables.

\subsection{Paramagnetic phase}
\begin{center}
\begin{figure}[!ht]
        \centerline{\psfig{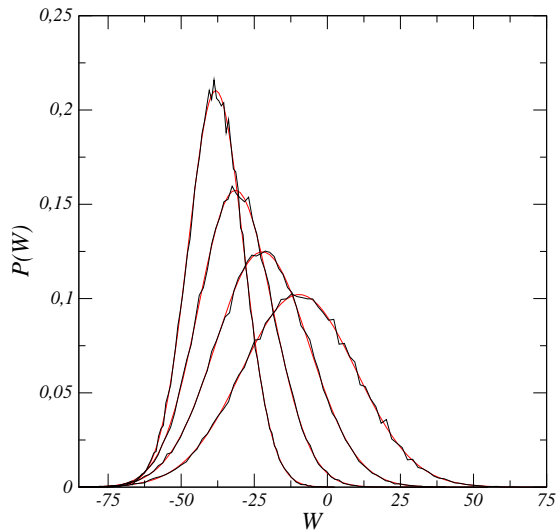}}
        \caption{Probability distribution of the work in the paramagnetic phase
	($\beta=0.2$) when applying linearly a magnetic field $h=0.1$ at different
	rates: $n_{\rm iter.}=2,5,10$ and $20$ (from right to left). The monotonous
	curves correspond to Gaussian fits.}
\label{fig1}
\end{figure}
\end{center}

As the temperature is increased in the paramagnetic phase, the correlation
length becomes smaller and eventually gets smaller than the lattice spacing. One
can thus consider the system as a set of free spins. The dynamical evolution of
the probability distribution of each of them is governed by the Glauber master
equation~\cite{Glauber63}. Since individual spins relax very rapidly to
equilibrium, the probability distribution of the work $\wp(W)$ is expected to
be Gaussian~\cite{Marathe05}. It is indeed what is observed numerically for
$\beta=0.2$, as can be seen in Figure~\ref{fig1}. We have checked that there is
no observable deviation from a Gaussian law in the tails. Calculation of the coefficient
of excess $\gamma=\langle (W-\langle W\rangle)^4\rangle/\langle (W-\langle
W\rangle)^2\rangle^2-3$ gives small values between $3.10^{-3}$ and $10^{-2}$ for
the different simulations performed.
\\

\begin{table}[!ht]
\begin{center}
\begin{tabular}{@{}*{6}{l}}
$h$ & $n_{\rm iter.}$ & $\langle W\rangle$ & $\Delta F$ (Jarzynski) & 
$W_{\rm diss.}=\langle W\rangle-\Delta F$ & $\Delta F$ (Gaussian approx.) \\
\hline
0.1 & 2 &
$-8.83(6)$ & 
$-4.5(1).10^{1}$ & 
$3.6(1).10^{1}$ & 
$-4.69(3).10^{1}$  
\\
0.1 & 5 &
$-2.155(5).10^{1}$ & 
$-4.70(10).10^{1}$ & 
$2.5(1).10^{1}$ & 
$-4.70(3).10^{1}$  
\\
0.1 & 10 &
$-3.068(4).10^{1}$ & 
$-4.8(1).10^{1}$ & 
$1.7(1).10^{1}$ & 
$-4.67(3).10^{1}$  
\\
0.1 & 15 &
$-3.523(3).10^{1}$ & 
$-4.68(1).10^{1}$ & 
$1.16(2).10^{1}$ & 
$-4.68(3).10^{1}$  
\\
0.1 & 20 &
$-3.777(3).10^{1}$ & 
$-4.678(9).10^{1}$ & 
$9.0(1)$ & 
$-4.68(3).10^{1}$  
\\
0.1 & 30 &
$-4.059(3).10^{1}$ & 
$-4.683(5).10^{1}$ & 
$6.23(8)$ & 
$-4.69(2).10^{1}$ 
\\
\hline
1.0 & 2 &
$-8.578(6).10^{2}$ & 
$-1.667(5).10^{3}$ & 
$3.7(1).10^3\ \dagger$ & 
$-4.6(1).10^{3}$  
\\
1.0 & 5 &
$-2.0500(5).10^{3}$ & 
$-2.616(2).10^{3}$ & 
$2.4(2).10^{3}\ \dagger$ & 
$-4.5(2).10^{3}$  
\\
1.0 & 10 &
$-2.9068(4).10^{3}$ & 
$-3.523(5).10^{3}$ & 
$1.5(2).10^{3}\ \dagger$ & 
$-4.4(2).10^{3}$  
\\
1.0 & 15 &
$-3.3037(3).10^{3}$ & 
$-3.711(5).10^{3}$ & 
$1.0(2).10^{3}\ \dagger$ & 
$-4.3(2).10^{3}$  
\\
1.0 & 20 &
$-3.5270(3).10^{3}$ & 
$-3.864(4).10^{3}$ & 
$0.8(2).10^{3}\ \dagger$ & 
$-4.3(2).10^{3}$  
\\
1.0 & 30 &
$-3.7663(2).10^{3}$ & 
$-4.019(3).10^{3}$ & 
$0.5(2).10^{3}\ \dagger$ & 
$-4.3(2).10^{3}$  
\\
\hline
10. & 2 &
$-4.5445(5).10^{4}$ & 
$-5.2532(5).10^{4}$ & 
- & 
$-3.0(5).10^{5}$  
\\
10. & 5 &
$-9.1721(3).10^{4}$ & 
$-9.5794(5).10^{4}$ & 
- & 
$-1.9(6).10^{5}$  
\\
10. & 10 &
$-1.12169(2).10^{5}$ & 
$-1.14980(5).10^{5}$ & 
- & 
$-1.5(5).10^{5}$  
\\
10. & 15 &
$-1.19695(2).10^{5}$ & 
$-1.21706(4).10^{5}$ & 
- & 
$-1.5(4).10^{5}$  
\\
10. & 20 &
$-1.23634(1).10^{5}$ & 
$-1.25357(5).10^{5}$ & 
- & 
$-1.4(3).10^{5}$  
\\
10. & 30 &
$-1.27737(1).10^{5}$ & 
$-1.29189(5).10^{5}$ & 
- & 
$-1.4(3).10^{5}$  
\\
\hline
\end{tabular}
\end{center}
\caption{Estimates of the average work $\langle W\rangle$, the free energy
difference $\Delta F$, the dissipated work $\langle W\rangle-\Delta F$
and the free energy difference using the Gaussian approximation in the
paramagnetic phase ($\beta=0.2$) for different magnetic fields $h$ and
transformation rates. The estimates of the dissipated work $W_{\rm diss.}$
marked with $\dagger$ have been computed using $\Delta F$ as given by the Gaussian
approximation because the Jarzynski equality fails in this case to give
stable values. In the case $h=10$, we do not even give any estimate
since the Gaussian approximation leads to errors bars larger than the estimate
of $W_{\rm diss.}$.}
\label{Table1}
\end{table}

The free-energy $\Delta F$ can be measured using the Jarzynski equality
(\ref{eq1}), either directly or by using the assumption that the work $W$ is
Gaussian distributed:
	\begin{equation}
	\langle e^{-\beta W}\rangle={1\over\sqrt{2\pi\sigma_W^2}}
	\int e^{-\beta W-(W-\langle W\rangle)^2/2\sigma_W^2}dW
	\Leftrightarrow\ \Delta F=\langle W\rangle-{\sigma_W^2\over 2k_BT}.
	\label{eq2}
	\end{equation}
This last relation was first obtained by Hermans~\cite{Hermans91}. In both
methods, the free energy difference should not depend on the experimental
protocol, in our case the rate at which the magnetic field is grown.
Our numerical results are summarized in Table~\ref{Table1}. For a small magnetic
field $h=0.1$ (to be compared with $J=1$), the two methods give, as expected,
estimates of $\Delta F$ in agreement independently of $n_{\rm iter.}$, i.e. of
the rate at which the transformation is performed. Using the Jarzynski relation
(\ref{eq1}), the slower the transformation, i.e. the more reversible, and the
more accurate the estimates are. Due to the amplification with the factor
${\rm exp}(-\beta W)$, the relevant information is in the negative tail of the
distribution. Insufficient sampling of this tail may lead to deviations of the
estimate of $\Delta F$. However, these deviations remain relatively small in our
system, as can be seen in Table~\ref{Table2}. The worst case is $\Delta
F=-40(2)$ obtained for $n_{\rm iter.}=2$ and $n_{\rm exp.}=1562$ while
$\Delta F\simeq -46.8$ for a larger statistics. The Gaussian assumption
(\ref{eq2}) turns out to lead to estimates of $\Delta F$ less noisy than
equation (\ref{eq1}) and that do not seem to depend on $n_{\rm iter.}$ (see
Table~\ref{Table1}). The relevant information comes in this case from the more
probable part of the distribution. However, as can be seen in
Table~\ref{Table2}, the estimate gets noisier faster for low statistics than
using the Jarzynski relation (\ref{eq1}). We also checked that the free energy
differences are extensive (Table~\ref{Table3}). When intermediate magnetic fields
$h=J$ are applied, the Jarzynski equality fails to give estimates of $\Delta F$
independent of the transformation rate while it is still the case for the
Gaussian approximation. When larger magnetic fields are applied, the two methods
fail to give estimates of $\Delta F$ for large transformation rates, i.e.
$n_{\rm iter.}$ small. A much larger statistics would be then required. As
well-known in the context of thermodynamic perturbation, one can circumvent
the problem by dividing the interval $[0;h]$ in smaller pieces and resorting to
several simulations to estimate $\Delta F$ in each of them.
\\

\begin{table}[!ht]
\begin{center}
\begin{tabular}{@{}*{6}{l}}
$n_{\rm exp.}$ & $\langle W\rangle$ & $\Delta F$ (Jarzynski) &
$W_{\rm diss.}=\langle W\rangle-\Delta F$ & $\Delta F$ (Gaussian approx.) \\
\hline
1562 &
$-3.80(2).10^{1}$ & 
$-4.67(5).10^{1}$ & 
$8.8(7)$ & 
$-4.7(2).10^{1}$  
\\
3125 &
$-3.76(2).10^{1}$ & 
$-4.63(3).10^{1}$ & 
$8.7(5)$ & 
$-4.6(1).10^{1}$  
\\
6250 &
$-3.78(1).10^{1}$ & 
$-4.62(2).10^{1}$ & 
$8.4(3)$ & 
$-4.6(1).10^{1}$  
\\
12,500 &
$-3.765(8).10^{1}$ & 
$-4.63(2).10^{1}$ & 
$8.7(3)$ & 
$-4.65(7).10^{1}$  
\\
25,000 &
$-3.773(6).10^{1}$ & 
$-4.67(2).10^{1}$ & 
$9.0(2)$ & 
$-4.67(5).10^{1}$  
\\
50,000 &
$-3.777(4).10^{1}$ & 
$-4.68(1).10^{1}$ & 
$9.0(2)$ & 
$-4.68(4).10^{1}$  
\\
100,000 &
$-3.777(3).10^{1}$ & 
$-4.678(9).10^{1}$ & 
$9.0(1)$ & 
$-4.68(3).10^{1}$  
\\
\hline
\end{tabular}
\end{center}
\caption{Estimates of the average work $\langle W\rangle$, the free energy
difference $\Delta F$, the dissipated work $\langle W\rangle-\Delta F$
and the free energy difference using the Gaussian approximation in the
paramagnetic phase $\beta=0.2$ with $h=0.1$ and $n_{\rm iter.}=20$
versus the number of experiments $n_{\rm exp.}$ used for the calculation of
the averages.}
\label{Table2}
\end{table}

\begin{table}[!ht]
\begin{center}
\begin{tabular}{@{}*{6}{l}}
$L$ & $n_{\rm iter.}$ & $\Delta F/L^2$ (Jarzynski) & $\Delta F/L^2$ (Gaussian approx.) \\
\hline
32 & 2 &
$-2.86(2).10^{-3}$ & 
$-2.87(3).10^{-3}$ \\ 
64 & 2 &
$-2.87(4).10^{-3}$ & 
$-2.86(2).10^{-3}$ \\ 
128 & 2 &
$-2.73(6).10^{-3}$ & 
$-2.86(2).10^{-3}$ \\ 
\hline
32 & 5 &
$-2.85(1).10^{-3}$ & 
$-2.84(2).10^{-3}$ \\ 
64 & 5 &
$-2.85(1).10^{-3}$ & 
$-2.84(2).10^{-3}$ \\ 
128 & 5 &
$-2.87(6).10^{-3}$ & 
$-2.87(2).10^{-3}$ \\ 
\hline
32 & 10 &
$-2.85(1).10^{-3}$ & 
$-2.86(2).10^{-3}$ \\ 
64 & 10 &
$-2.849(7).10^{-3}$ & 
$-2.85(1).10^{-3}$ \\ 
128 & 10 &
$-2.94(7).10^{-3}$ & 
$-2.85(2).10^{-3}$ \\ 
\hline
32 & 20 &
$-2.865(8).10^{-3}$ & 
$-2.87(1).10^{-3}$ \\ 
64 & 20 &
$-2.850(5).10^{-3}$ & 
$-2.85(1).10^{-3}$ \\ 
128 & 20 &
$-2.855(5).10^{-3}$ & 
$-2.85(2).10^{-3}$ \\ 
\hline
\end{tabular}
\end{center}
\caption{Estimates of the free energy difference $\Delta F$ in the
paramagnetic phase ($\beta=0.2$ and $h=0.1$) for different lattice sizes $L$
and transformation rates $h/n_{\rm iter.}$.}
\label{Table3}
\end{table}

\begin{center}
\begin{figure}[!ht]
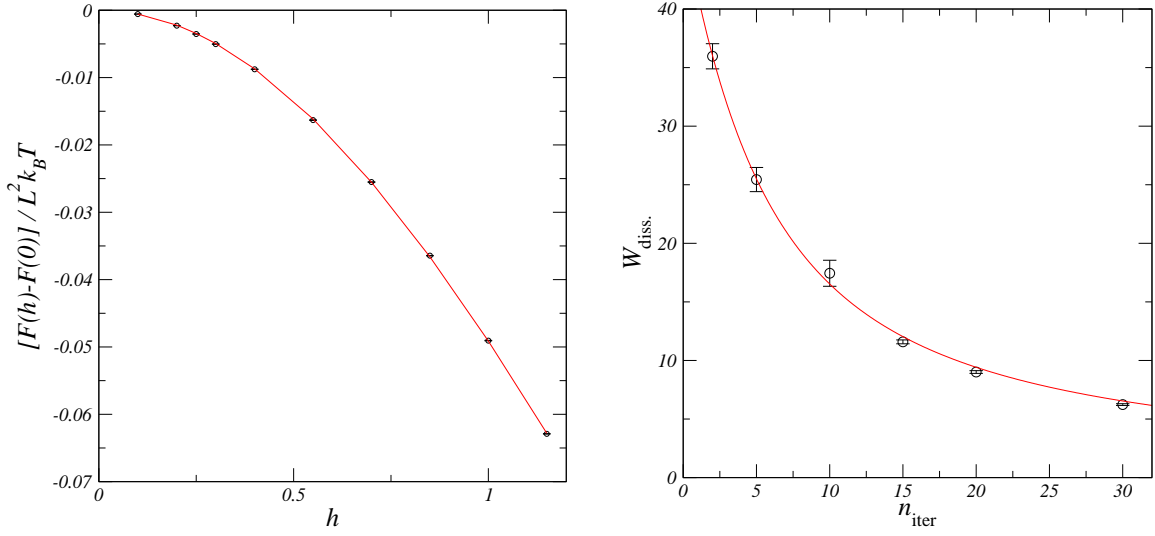

        \centerline{\psfig{figure=dF-Para.eps,height=7cm}\quad\quad
	\psfig{figure=Wdiss-Para.eps,height=7cm}}
        \caption{On the left, free energy density $\Delta F/L^2$ versus the final
	applied	magnetic field $h$ in the paramagnetic phase ($\beta=0.2$) for
	the slowest transformation rate considered ($n_{\rm iter.}=30$). The
	curve corresponds to a non-linear fit with the expression (\ref{ExprdF}).
	On the right, dissipated work $W_{\rm diss.}$ with respect to the number of
	Monte Carlo iterations $n_{\rm iter.}$ for $h=0.1$. The curve is a
	two-parameter non-linear fit with the expression (\ref{ExprWdiss}).}
\label{fig2}
\end{figure}
\end{center}

The results presented above can be understood by assuming that the system
behaves as $N$ independent Ising domains with an average magnetic moment $m$.
The free energy difference between the equilibrium states with and without
magnetic field $h$ reads
	\begin{equation}
	\Delta F=F(h)-F(0)=-Nk_BT\ln\cosh{mh\over k_BT}
	\label{ExprdF}
	\end{equation}
The numerical data are in very good agreement with this law as can be seen in
Figure~\ref{fig2}. The best fit is obtained for the parameters $N\simeq 0.13L^2$
and $m\simeq 4.62$ which means that the characteristic size of the domains is
of the order of $\xi\sim\sqrt{L^2/N}\simeq 2.8$ and that the domains are not
at saturation, $Nm/L^2\simeq 0.6$. At equilibrium, the magnetization is
expected to be
	\begin{equation}
	M_{\rm eq.}(h)=-{\partial F\over\partial h}=Nm\tanh{mh\over k_BT}
	\simeq N{m^2h\over k_BT}
	\label{MChpMoyen}
	\end{equation}
Let us assume the phenomenological evolution equation
	\begin{equation}
	{\partial M\over\partial t}=-{M(t)-M_{\rm eq.}(h)\over\tau}
	\ \Leftrightarrow\ M(t)=M(0)e^{-t/\tau}
	+{1\over\tau}\int_0^t M_{\rm eq.}(h(t'))e^{-(t-t')/\tau}dt'
	\label{EqMvtPara}
	\end{equation}
where $\tau$ is the relaxation time. The work extracted from the system
when the magnetic field is increased linearly, i.e. $h(t)=\dot ht$, reads
	\begin{eqnarray}
	W=-\dot h\int M(t')dt'
	&=&-\dot hM(0)\big(1- e^{-t/\tau}\big)-{\dot h\over\tau}
	\int_0^t\int_0^{t'}M_{\rm eq.}(h(t''))e^{-(t'-t'')/\tau}dt'dt''
	\nonumber \\
	&\simeq &-\dot hM(0)\big(1- e^{-t/\tau}\big)-{\dot h^2Nm^2\over\tau k_BT}
	\int_0^t\int_0^{t'} t''e^{-(t'-t'')/\tau}dt'dt''		\\
	&=&-\dot hM(0)\big(1- e^{-t/\tau}\big)-{\dot h^2Nm^2\over k_BT}
	\left[{1\over 2}t^2-\tau t+\tau^2\left(1-e^{-t/\tau}\right)\right]
	\nonumber
	\end{eqnarray}
where $M_{\rm eq.}$ has been replaced by its weak-field expansion
(\ref{MChpMoyen}). Starting the experiment from the paramagnetic phase, i.e.
$M(0)=0$, and subtracting the reversible work
	\begin{equation}
	W_{\rm rev.}=-\dot h\int M_{\rm eq.}(t')dt'
	\simeq -{\dot h^2Nm^2\over 2k_BT}t^2
	\end{equation}
one gets finally	
	\begin{equation}
	W_{\rm diss.}\simeq {\dot h^2Nm^2\over k_BT}
	\left[\tau t-\tau^2\left(1-e^{-t/\tau}\right)\right]
	=\dot hM_{\rm eq.}(h)\tau\left[1-{\tau\over t}\left(1-e^{-t/\tau}
	\right)\right]
	\label{ExprWdiss}
	\end{equation}
As can be seen in Figure \ref{fig2}, the expression fits well the numerical
data for $h=0.1$. The non-linear fit gives $\tau\simeq 2.23$ for the relaxation
time and $m\simeq 4.72 $ assuming $N=0.13L^2$. Note that the estimate
$m$ is compatible with the one calculated from $\Delta F$ ($m\simeq 4.62$).
The estimate of the relaxation time $\tau$ is small as expected for
such a high temperature. For larger magnetic fields, the dissipated work is
still nicely fitted by the expression (\ref{ExprWdiss}) but the fit gives
too small average moments $m$. This is due to the fact that the expression
(\ref{ExprWdiss}) was derived in the weak magnetic field limit.

\subsection{Ferromagnetic phase}
In the ferromagnetic phase $\beta=0.7$, the probability distributions $\wp(W)$
display two well-separated peaks corresponding to the work performed by systems
in each one of the two initial ferromagnetic ground states. The average $\langle
e^{-\beta W}\rangle$ is dominated by the peak at more negative values of the
work. As a consequence, the Gaussian approximation applied only to this peak
leads to estimates of $\Delta F$ in very good agreement with those obtained
with the Jarzynski equality (Table~\ref{Table4}). Since the two peaks are not
centered around zero, averages over the most negative peak were computed in
practise by selecting configurations for which the work $W$ is smaller than the
average of the smallest and largest works. A large gap separating the two peaks,
the average work would have given the same results. As can be seen in
Table~\ref{Table4}, the estimates of $\Delta F$ do not show any dependence on
the transformation rate and give compatible values using the Jarzynski equation
(\ref{eq1}) and the Gaussian approximation (\ref{eq2}). 
\\

\begin{table}[!ht]
\begin{center}
\begin{tabular}{@{}*{6}{l}}
$h$ & $n_{\rm iter.}$ & $\langle W\rangle$ & $\Delta F$ (Jarzynski) &
$W_{\rm diss.}=\langle W\rangle-\Delta F$ & $\Delta F$ (Gaussian approx.) \\
\hline
0.1 & 10 &
$1.0(5).10^{1}$ & 
$-1.622740(7).10^{3}$ & 
$1.633(5).10^{3}$ & 
$-1.624(5).10^{3}$  
\\
0.1 & 50 &
$7(5)$ & 
$-1.622732(5).10^{3}$ & 
$1.629(5).10^{3}$ & 
$-1.624(2).10^{3}$  
\\
0.1 & 250 &
$4(5)$ & 
$-1.622735(5).10^{3}$ & 
$1.627(5).10^{3}$ & 
$-1.624(1).10^{3}$  
\\
\hline
1.0 & 10 &
$-2.4(5).10^{2}$ & 
$-1.63057(3).10^{4}$ & 
$1.607(5).10^{4}$ & 
$-1.63(4).10^{4}$  
\\
1.0 & 50 &
$-1.40(5).10^{3}$ & 
$-1.630612(4).10^{4}$ & 
$1.490(5).10^{4}$ & 
$-1.63(2).10^{4}$  
\\
1.0 & 250 &
$-6.22(3).10^{3}$ & 
$-1.6306156(9).10^{4}$ & 
$1.009(3).10^{4}$ & 
$-1.631(7).10^{4}$  
\\
\hline
10. & 10 &
$-1.076(2).10^{5}$ & 
$-1.63676(1).10^{5}$ & 
$5.61(2).10^{4}$ & 
$-1.6(2).10^{5}$  
\\
10. & 50 &
$-1.3313(10).10^{5}$ & 
$-1.637355(8).10^{5}$ & 
$3.061(10).10^{4}$ & 
$-1.64(7).10^{5}$  
\\
10. & 250 &
$-1.4475(6).10^{5}$ & 
$-1.637402(2).10^{5}$ & 
$1.899(6).10^{4}$ & 
$-1.64(3).10^{5}$  
\\
\hline
\end{tabular}
\end{center}
\caption{Estimates of the average work $\langle W\rangle$, the free energy
difference $\Delta F$, the dissipated work $\langle W\rangle-\Delta F$
and the free energy difference using the Gaussian approximation with the most
negative peak in the ferromagnetic phase ($\beta=0.7$) for different magnetic
fields $h$ and transformation rates $h/n_{\rm iter.}$.}
\label{Table4}
\end{table}

In a sense, the physics is simpler than in the paramagnetic phase. Before
applying the magnetic field, the initial magnetization is either $+M_0$ or
$-M_0$ with small fluctuations around these two values. When applying a small
magnetic field, i.e. $h=0.1$ in our case, the magnetization does not change
much so that the work is either $-M_0h$ or $+M_0h$. According to Jarzynski
equation (\ref{eq1}), the free energy reads
	\begin{equation}
	\Delta F=-k_BT\ln\langle e^{-\beta W}\rangle
	=-k_BT\ln\left[{1\over 2}e^{-\beta M_0h}+{1\over 2}e^{\beta M_0h}\right]
	\simeq -M_0h+k_BT\ln 2
	\end{equation}
Using the saturation magnetization $M_{\rm sat}=L^2=16384$, one obtains indeed
a good estimate of $\Delta F$. The average work is small $\langle
W\rangle=(-M_0h+M_0h)/2=0$ so that the dissipated work is $W_{\rm
diss.}=\langle W\rangle-\Delta F\simeq -\Delta F$. For larger magnetic fields,
$h=1$ and $h=10$, this approximation is not valid anymore: the magnetization
tends to reverse itself if it was initially in the opposite direction of the
field. However, since the Jarzynski equality (\ref{eq1}) is dominated by the
systems for which the magnetization was initially in the same direction than the
field, one still expects that $\Delta F\simeq -M_{\rm sat}h$. But the average
magnetization now moves towards $-M_{\rm sat}$ so $W_{\rm diss.}>0$. Obviously
dissipation is thus mainly due to the systems for which the initial magnetization
was in the ``wrong'' direction.

\subsection{Critical point}
We now present results at the critical point $\beta_c={1\over 2}\ln\left(1+
\sqrt 2\right)\simeq 0.440687$. Like in the ferromagnetic phase, the probability
distributions $\wp(W)$ display two well-separated peaks. Due to finite-size
effects, the initial state has indeed a non-vanishing magnetization of sign
either positive or negative due to the $Z_2$ symmetry ($\sigma_i\rightarrow
-\sigma_i$) of the Ising model. In contradistinction to the ferromagnetic case,
the two peaks do not have a Gaussian shape. The free energy differences
$\Delta F$ calculated from the Jarzynski equation (\ref{eq1}) are presented for
two different magnetic fields in Table~\ref{Table5}. For the smallest magnetic
field $h\simeq 0.023$, the free-energy turns out to be independent of the
transformation rate $h/n_{\rm iter.}$: the maximum deviation from the average
is of the order of one standard deviation. For the largest magnetic field
$h\simeq 0.23$, a systematic deviation is observed.

\begin{table}[!ht]
\begin{center}
\begin{tabular}{@{}*{6}{l}}
$h$ & $n_{\rm iter.}$ & $\langle W\rangle$ & $\Delta F$ (Jarzynski) &
$W_{\rm diss.}=\langle W\rangle-\Delta F$ \\
\hline
0.022692 & 2 &
$-1.6(7)$ & 
$-2.690(7).10^{2}$ & 
$2.67(1).10^{2}$  
\\
0.022692 & 10 &
$-1.1(7)$ & 
$-2.697(4).10^{2}$ & 
$2.69(1).10^{2}$  
\\
0.022692 & 50 &
$-7.5(7)$ & 
$-2.704(5).10^{2}$ & 
$2.63(1).10^{2}$  
\\
0.022692 & 100 &
$-1.39(7).10^{1}$ & 
$-2.701(4).10^{2}$ & 
$2.56(1).10^{2}$  
\\
0.022692 & 150 &
$-1.85(7).10^{1}$ & 
$-2.698(2).10^{2}$ & 
$2.513(8).10^{2}$  
\\
0.022692 & 250 &
$-2.87(7).10^{1}$ & 
$-2.699(1).10^{2}$ & 
$2.411(8).10^{2}$  
\\
\hline
0.226918 & 2 &
$-2.4(7).10^{1}$ & 
$-2.897(2).10^{3}$ & 
$2.873(9).10^{3}$  
\\
0.226918 & 10 &
$-1.87(7).10^{2}$ & 
$-3.009(2).10^{3}$ & 
$2.822(9).10^{3}$  
\\
0.226918 & 50 &
$-7.93(6).10^{2}$ & 
$-3.103(1).10^{3}$ & 
$2.310(7).10^{3}$  
\\
0.226918 & 100 &
$-1.362(5).10^{3}$ & 
$-3.133(2).10^{3}$ & 
$1.771(6).10^{3}$  
\\
0.226918 & 150 &
$-1.680(4).10^{3}$ & 
$-3.136(2).10^{3}$ & 
$1.456(6).10^{3}$  
\\
0.226918 & 250 &
$-2.017(3).10^{3}$ & 
$-3.143(1).10^{3}$ & 
$1.126(4).10^{3}$  
\\
\hline
\end{tabular}
\end{center}
\caption{Estimates of the average work $\langle W\rangle$, the free energy
difference $\Delta F$, the dissipated work $\langle W\rangle-\Delta F$
at the critical point for two different transformation rates
$h/n_{\rm iter.}$ and two magnetic fields $h$.}
\label{Table5}
\end{table}

We made simulations for ten values of the final magnetic field in the range
$h\in[0.023;0.23]$. The singular part of the free-energy is expected to scale at
the critical point as
	\begin{equation}
	F_{\rm sing.}(h)\build\sim_{T\rightarrow T_c,h\ll 1}^{} h^{1+1/\delta}
	\label{Scaling}
	\end{equation}
Neglecting the contribution of the regular part, we fitted $\Delta F$ as given
by the Jarzynski equality with the scaling form (\ref{Scaling}). The data display
a nice power-law behavior but fluctuations are observed for fast transformations
($n_{\rm iter.}\le 10$) at large magnetic fields. The estimates of $1+1/\delta$
and $\delta$ are collected in Table~\ref{Table6}. As the transformation
gets slower, and thus $n_{\rm iter.}$ larger, the numerical estimate gets
closer to the exact result $\delta=15$. For the fastest transformation
$n_{\rm iter.}=2$, the relative deviation of the numerical estimate from the
exact result is of order of 100\%! Note however that much better estimates are
obtained when restricting the fit to the smallest values of the magnetic field:
$\delta=19.9(7)$ for $h\le 0.018$ and $\delta=14.1(6)$ for $h\le 0.013$
when $n_{\rm iter.}=2$. Since error bars take into account all sources of
statistical fluctuations, the deviation may be explained as a systematic bias
due to a too small number of experiments for the calculation of the average.
As can be seen in Table~\ref{Table7}, when $n_{\rm exp.}$ is made smaller
the estimate of $\delta$ indeed increases. The effect is dramatic for fast
transformations: a relatively good estimate of $\delta$ is already obtained for
$n_{\rm iter.}=250$ when averaging over only $n_{\rm exp.}=1000$ experiments
but the same deviation from the exact result $\delta=15$ is not even
obtained with $n_{\rm exp.}=100,000$ experiments for $n_{\rm iter.}=50$.
More stable estimates of $\delta$ are obtained for a smaller system, $L=64$, for
which $\delta$ does not display a systematic deviation but fluctuates between $14.0(1)$
($n_{iter}=2$) and $14.1(2)$ ($n_{iter}=150$) and remains below the exact value
because of finite-size effects. 

\begin{table}[!ht]
\begin{center}
\begin{tabular}{@{}*{6}{l}}
$n_{\rm iter.}$ & $1+1/\delta$ & $\delta$ \\
\hline
$2$ & $1.024(7)$ & $42(1)$ \\ 
$10$ & $1.045(6)$ & $22.2(3)$ \\ 
$50$ & $1.059(5)$ & $17.0(1)$ \\ 
$100$ & $1.064(4)$ & $15.7(1)$ \\ 
$150$ & $1.065(3)$ & $15.36(7)$ \\ 
$250$ & $1.066(2)$ & $15.13(5)$ \\ 
\hline
\end{tabular}
\end{center}
\caption{Critical exponents $1+1/\delta$ and $\delta$ obtained by power-law interpolation
of the free energy $\Delta F=F(h)-F(0)\sim h^{1+1/\delta}$ with ten magnetic fields
in the range $[0.023;0.23]$ at the critical point for different transformation rates
$h/n_{\rm iter.}$.}
\label{Table6}
\end{table}

\begin{table}[!ht]
\begin{center}
\begin{tabular}{@{}*{6}{l}}
$n_{\rm exp.}$ & $1+1/\delta$ & $\delta$ \\
\hline
$10^2$ & $1.060(10)$ & $16.6(3)$ \\ 
$10^3$ & $1.066(7)$ & $15.1(2)$ \\ 
$10^4$ & $1.065(3)$ & $15.44(8)$ \\ 
$10^5$ & $1.066(2)$ & $15.13(5)$ \\ 
\hline
\end{tabular}
\end{center}
\caption{Critical exponents $1+1/\delta$ and $\delta$ obtained by power-law interpolation
of the free energy $\Delta F=F(h)-F(0)\sim h^{1+1/\delta}$ with ten magnetic fields
in the range $[0.023;0.23]$ at the critical point for different number of experiments
$n_{\rm exp.}$ in the slowest case $n_{\rm iter.}=250$.}
\label{Table7}
\end{table}

\begin{center}
\begin{figure}[!ht]
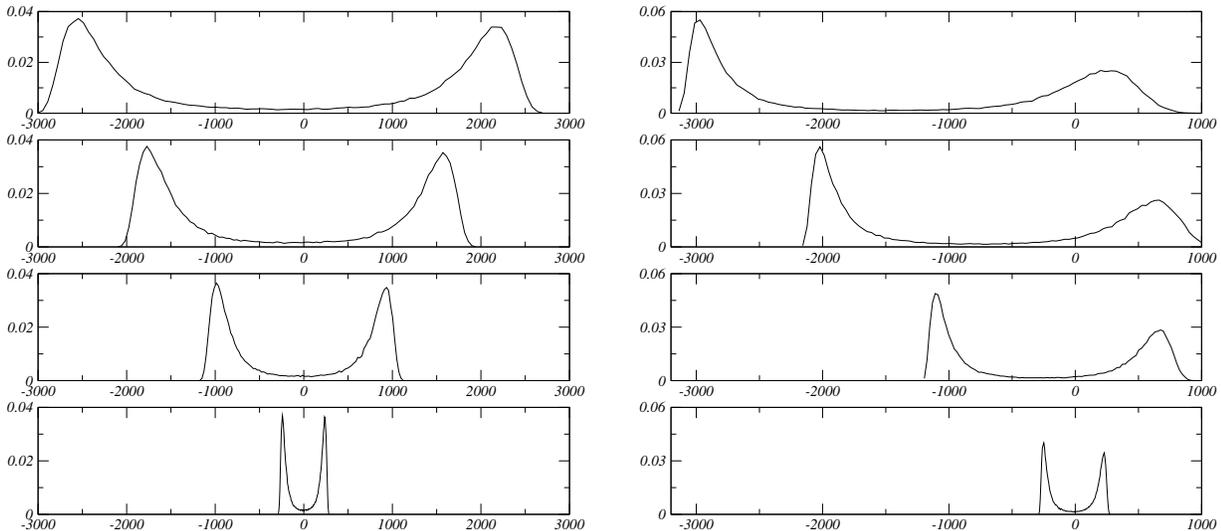

        \centerline{\psfig{figure=Proba-Tc-Niter=10.eps,height=7cm}\quad\quad
	\psfig{figure=Proba-Tc-Niter=100.eps,height=7cm}}
        \caption{On the left, probability distributions $\wp(W)$ versus the work $W$
	for a fast transformation $n_{\rm iter.}=10$ for four different final magnetic
	fields $h=0.004, 0.017, 0.031$ and $0.044$ (from bottom to top).
	On the right, probability distribution $\wp(W)$ versus $W$ for an intermediate
	transformation $n_{\rm iter.}=100$ for the same values of the magnetic field
	(from bottom to top).}
\label{fig3}
\end{figure}
\end{center}

\begin{center}
\begin{figure}[!ht]
        \centerline{\psfig{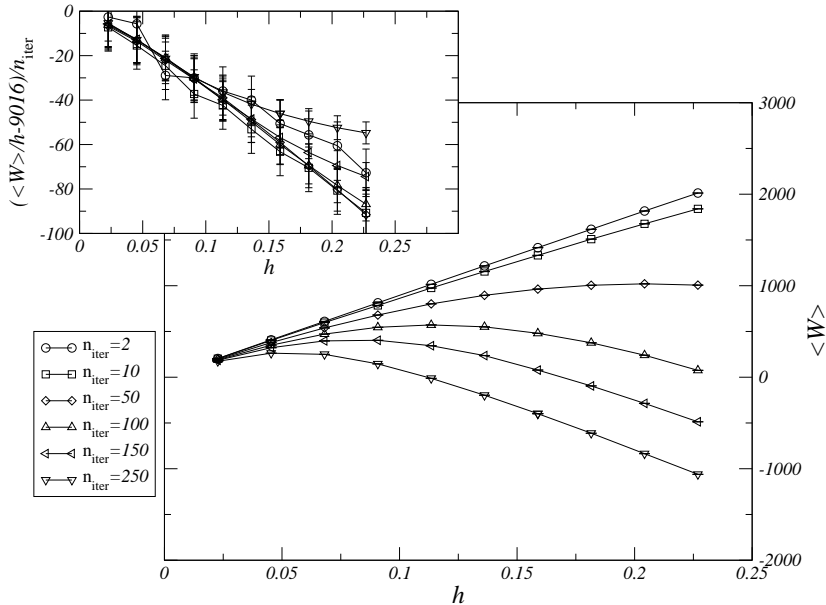}}
        \caption{Average work $\langle W\rangle$ of the right peak of the probability
	distribution at the critical point versus the final magnetic field $h$
	for different transformation rates, i.e. $n_{\rm iter.}$ from 2 to 250.
	In the inset, $\big(\langle W\rangle/h-M_0\big)/n_{iter.}$ where $M_0$ has been
	adjusted to the value $M_0=9016$ is plotted versus $h$ to check the linear
	behavior predicted by (\ref{WorkTc}).}
\label{fig4}
\end{figure}
\end{center}

As already mentioned, the two-peak structure of the probability distribution of the
work $\wp(W)$ differs from the ferromagnetic case. As can be seen in Figure~\ref{fig3},
the left peak, which dominates the average in the Jarzynski equality (\ref{eq1}) and
thus the free energy difference $\Delta F$, moves monotonously to the left as the
magnetic field is increased. The right peak moves first to the right, up to a certain
magnetic field before pursing to the left. This behavior can be understood by assuming
that the dynamics of the magnetization is governed by the Langevin equation
	\begin{equation}
	{\partial M\over\partial t}=-{\delta{\cal F}\over\delta M}\simeq \kappa h(t)
	\ \Leftrightarrow\ M(t)\simeq M(0)+\kappa\int_0^t h(t')dt'
	\label{eqLangevin}
	\end{equation}
where the link with equation (\ref{EqMvtPara}) is made by considering $\kappa$ as a
magnetic susceptibility divided by the relaxation time $\tau$. It is sufficient for
the discussion to consider $\kappa$ as constant. For a single realization, the initial
magnetization $M(0)$ is non-zero due to finite-size effects. When the field is increased
linearly, i.e. $h(t)=\dot ht$, the work extracted from the system up to time $t$ is
	\begin{equation}
	W=-\int_0^t M(t)\dot hdt=-M(0)\dot ht-{\kappa\over 6}\dot h^2t^3
	=-M(0)h-{\kappa\over 6}h^2t
	\label{WorkTc}
	\end{equation}
When $M(0)<0$, the work initially increases and eventually at $h_t=3|M(0)|/\kappa t$ starts
to decrease as can be observed in Figures~\ref{fig3} and~\ref{fig4}. In our case, no turning
point is observed for $n_{\rm iter.}=10$ because the magnetic fields remain too small. The data
presented in the inset of figure~\ref{fig4} reproduce the linear behavior of $\big(W/h-W_0)/n_{\rm iter.}$
versus $h$ predicted by equation (\ref{WorkTc}). Higher order terms are observed only for
the slowest transformations. When the initial magnetization is initially in the same direction
as the magnetic field, i.e. $M(0)>0$, the work is always negative. For very slow transformation
rate, the magnetization is expected to follow the magnetic field, i.e. to behave as the
equilibrium magnetization $M_{\rm eq.}\sim h^{1/\delta}$. Taking into account the magnetic
field dependence of the $\kappa$, one should expect in this case the solution of the Langevin
equation (\ref{eqLangevin}) to scale as
	\begin{equation}
	W_{\rm rev.}\sim -\int_0^h {h'}^{1/\delta}dh'\sim -h^{1+{1\over\delta}}
	 \end{equation}
This behavior is indeed observed: a power-law interpolation of the position of the left
peak versus the magnetic field leads to exponents $\delta=14(1)$ for $n_{\rm iter}=150$
and $250$ compatible with the exact result $\delta=15$.
	
\section{Conclusions}
The Jarzynski equality offers the possibility to estimate free energy differences
in a very simple manner in Monte Carlo simulations. Despite the fact that this
relation is exact in the case of a Markovian dynamics, it has to be used carefully.
As already shown by several authors, we observed systematic deviations for strongly
irreversible transformations due to an insufficient sampling. These systematic deviations
are caused by the fact that the average $\langle e^{-\beta W}\rangle$ is dominated by
very rare events $-W\gg 1$ that may require a large number of simulations to be
properly sampled. This limitation of the Jarzynski equality is a strong one:
the error cannot be estimated simply. When the distribution of the work is Gaussian,
it has been shown that the number of experiments has to be grown exponentially with the
dissipated work $W_{\rm diss.}$~\cite{Ritort02}. We have observed that the Gaussian
approximation (\ref{eq2}) may still give reliable estimates of the free energy difference
$\Delta F$ when the Jarzynski equality fails as for example in the case of intermediate
magnetic fields applied on a paramagnet. Unfortunately, the Gaussian approximation is valid
only for one-particle systems~\cite{Mazonka99,Speck04} or systems with short-range
correlations like paramagnets but not at the critical point. However, the Jarzynski equality
turned out to be useful at small magnetic field. It was reliable enough for instance to
estimate the critical exponent $\delta$. For this purpose, the Jarzynski equality may be
useful since a single Monte Carlo simulation is required if the work is measured at
different times.

\section*{Acknowledgements}
The laboratoire de Physique des Mat\'eriaux is Unit\'e Mixte de Recherche
CNRS number 7556. The authors would like to thank the Statistical
Physics group and especially Olivier Collet for stimulating and illuminating
discussions and Bertrand Berche for critical reading of the paper.


\begin{thebibliography}{64}
\bibitem{Jarzynski97a} C. Jarzynski (1997) {\sl Phys. Rev. Lett.} {\bf 78}, 2690
\bibitem{Jarzynski97b} C. Jarzynski (1997) {\sl Phys. Rev. E} {\bf 56}, 5018.
\bibitem{Bochkov81}
G.N. Bochkov and Y.U. Kuzovlev (1981) {\sl Physica A} {\bf 106}, 443;
G.N. Bochkov and Y.U. Kuzovlev (1981) {\sl Physica A} {\bf 106}, 480
\bibitem{Crooks99} G.E. Crooks (1999) {\sl Phys. Rev. E} {\bf 60}, 2721
\bibitem{Liphardt02} J. Liphardt, S. Dumont, S.B. Smith, I. Tinoco and C. Bustamante
(2002) {\sl Science} {\bf 296}, 1832
\bibitem{Ritort03} F. Ritort (2003) Poincar\'e Seminar {\bf 2}, 195
\bibitem{Douarche05} F. Douarche, C. Ciliberto and A. Petrosyan (2005)
	{\sl J. Stat. Mech} P09011
\bibitem{Lua05} R.C. Lua and A.Y. Grosberg (2005)
	{\sl J. Chem. Phys. B} {\bf 109}, 6805
\bibitem{Cohen04} E.G.D. Cohen and D. Mauzerall (2004) {\sl J. Stat. Mech} P07006
\bibitem{Jarzynski04} C. Jarzynski (2004) {\sl J. Stat. Mech:Theor. Exp.} P09005 
\bibitem{Marathe05} Rahul Marathe and Abishek Dhar (2005) {\tt cond-mat/0508043}
\bibitem{Imparato05} A. Imparato and L. Peliti (2005)
	 Phys. Rev. E 72, 046114 (2005).
{\sl Europhys. Lett.}, {\bf 70}, 740
\bibitem{SwendsenWang87} R.H. Swendsen and J.S. Wang (1987) {\sl Phys. Rev. Lett.}
{\bf 58}, 86
\bibitem{Salas96} J. Salas and A.D. Sokal  (1996) {\sl J. Stat. Phys.} {\bf 85}, 297
\bibitem{Metropolis53} N. Metropolis, A.E. Rosenbluth, M.N. Rosenbluth,
A.H. Teller, E. Teller (1953) {\sl J. Chem. Phys.} {\bf 21}, 1087
\bibitem{Narayan03} O. Narayan and A. Dhar (2003) {\tt cond-mat}/0307148
\bibitem{Glauber63} R. Glauber (1963) {\sl J. Math. Phys.} {\bf 4}, 294
\bibitem{Hermans91} J. Hermans (1991) {\sl J. Chem. Phys.}, {\bf 95}, 9029
\bibitem{Ritort02} F. Ritort, C. Bustamante and I. Tinoco (2002) {\sl PNAS} {\bf 99},
13544
\bibitem{Mazonka99} O. Mazonka and C. Jarzynski (1999) {\tt cond-mat}/9912121
\bibitem{Speck04} T. Speck and U. Seifert (2004) {\sl Phys. Rev. E} {\bf 70}, 066112
\end{thebibliography}
\end{document}